\documentclass[conference, 10pt]{IEEEtran}
\IEEEoverridecommandlockouts
\usepackage{cite}
\usepackage{amsmath,amssymb,amsfonts}
\usepackage{algorithmic}
\usepackage{graphicx}
\usepackage{textcomp}
\usepackage{xcolor}
\usepackage{mathptmx}

\def\BibTeX{{\rm B\kern-.05em{\sc i\kern-.025em b}\kern-.08em
    T\kern-.1667em\lower.7ex\hbox{E}\kern-.125emX}}

\begin{document}

\title{Deep Learning Approach to Photometric Redshift Estimation\\

\thanks
}

\author{\IEEEauthorblockN{Krishna Chunduri}
\IEEEauthorblockA{\textit{College of CDSS} \\
University of California, Berkeley\\
CA, USA \\
krishnac24@berkeley.edu}
\and
\IEEEauthorblockN{Mithun Mahesh}
\IEEEauthorblockA{\textit{Department of Computer Science} \\
Purdue University\\
West Lafayette, IN, USA \\
mahesh70@purdue.edu}

}
\IEEEoverridecommandlockouts
\IEEEpubid{\makebox[\columnwidth]{978-1-6654-7345 3/22/\$31.00~\copyright~2024 IEEE \hfill}
\hspace{\columnsep}\makebox[\columnwidth]{ }}

\maketitle

\begin{abstract}
Data-driven approaches play a crucial role in space computing, and our paper focuses on analyzing data to learn more about celestial objects. Photometric redshift, a measure of the shift of light towards the red part of the spectrum, helps determine the distance of celestial objects. This study used a dataset from the Sloan Digital Sky Survey (SDSS) with five magnitudes alongside their corresponding redshift labels. Traditionally, redshift prediction relied on spectral distribution templates (SEDs), which, though effective, are costly and limited, especially for large datasets. This paper explores data-driven methodologies instead of SEDs. By employing a decision tree regressor and a fully connected neural network (FCN), we found that the FCN outperforms the decision tree regressor in RMS. The results show that data-driven estimation is a valuable tool for astronomical surveys. With the adaptability to complement previous methods, FCNs will reshape the field of redshift estimation.

\end{abstract}

\section{Introduction}
\subsection{Background}
Photometric redshift estimation is an essential process in modern astronomy, determining the redshift of celestial objects, such as galaxies, stars, and quasars. By measuring the object’s magnitude in different wavelength filters, such as ultraviolet (u) or green (g) and evaluating the differences in magnitude to determine the object’s color (u-g), we can use color values to help estimate redshift for the celestial object \cite{b1}. Such estimations play a crucial role in the interpretation and understanding of large astronomical data surveys, shedding light on distances for celestial objects. Acquiring accurate redshift data is imperative towards advancing our grasp on galaxy formation and evolution.

Traditional methods often employ spectroscopy to determine redshift, utilizing galaxy spectral signature and wavelength shifts. However, this technique can be resource-intensive and expensive. Furthermore, faint celestial objects can pose challenges to spectroscopic observations. These drawbacks have led to the emergence of photometric redshift as a viable alternative. Photometric redshift estimation harnesses the magnitude of extragalactic objects as observed across multiple filters \cite{b2}. Rather than relying on a detailed spectrum, astronomers utilize the intensity of light across select broad wavelength bands to infer redshift.

In the realm of galaxy evolution studies, the performance of photometric redshifts (photo-z’s) has profound implications. With systematic uncertainties in modeling galaxy evolution anticipated to persist in the foreseeable future, ensuring the precision of photometric redshift becomes even more important. For instance, the subdivision of objects according to their redshifts is instrumental in targeting specific redshift ranges in spectroscopic surveys. The overarching takeaway is clear: the efficacy of photo-z estimation is integral to the success of galaxy evolution studies. Creating a model that estimates photometric redshift given magnitude data is an optimal tool to assist many areas of research within the astronomical world. 

\subsection{Previous Research}
Previous studies have found significant advancements. The CANDELS GOODS-S survey, utilizing the Hubble Space Telescope (HST) Wide Field Camera 3 (WFC3) H-band and Advanced Camera for Surveys (ACS) z-band, has helped expand our understanding of photometric redshifts \cite{b3}. This dataset, with TFIT photometry, explored the efficacy of various codes and template Spectral Energy Distributions. It found that methods which incorporated training using a spectroscopic sample achieved enhanced accuracy. Importantly, the research found a direct correlation between the source magnitude and the precision of redshift estimation, emphasizing the role of magnitude in estimation.

Another approach was utilizing Bayesian methodologies \cite{b4}. By employing prior probabilities and Bayesian marginalization, this method was adept at utilizing previously overlooked data like the expected shape of redshift distributions and galaxy type fractions. When applied to B130 HDF-N spectroscopic redshifts, this Bayesian approach showcased promising results, reinforcing its potential to address existing gaps. Importantly, these advancements were realized without the reliance on a training-set procedure, while utilizing template libraries.

Both studies used template Spectral Energy Distribution (SED) data to help test their different methodologies. While template SEDs do help estimate photometric redshift, it’s become increasingly more difficult to obtain these distributions with larger datasets. Given the next generation of surveys from the James Webb Space Telescope (JWST) and Rubin Observatory (LSST), photometric redshift estimation needs a more data-driven approach to accurately predict redshift based on observational data \cite{b5}.

A few more studies highlighted approaches using machine learning to predict redshift using data from SDSS. One uses TPZ \cite{b6} which is a parallel machine learning algorithm that generates photometric redshift Probability Density Functions(PDFs) using prediction trees and random forest techniques, efficiently incorporating measurement errors and handling missing values. It provides supplementary data analysis information, including unbiased accuracy estimates, identification of poor photometric redshift areas, variable importance quantification, and robust outlier identification, enhancing the efficacy of redshift estimation. Another used artificial neural networks in a supervised learning approach to use data and corresponding labels to train their model, achieving an RMS of 0.023 \cite{b7}. Our model draws inspiration from these approaches as well but is different in methodology because of the nonstandard architecture we use and also improves the RMS to a lower value which indicates a higher accuracy during predictions. 

A separate approach addresses class-dependent residuals and mode collapse biases in CNN-based photometric redshift estimation, proposing methods like multi-channel outputs, balanced training data, and soft labels to mitigate these biases. The proposed methods demonstrate improved bias control and robustness, highlighting their potential for future cosmological surveys and other data-driven models \cite{b8}.

\subsection{Our Work}
The primary objective of this paper is to explore novel computational methods that take a data-driven approach to estimation while increasing accuracy. A data-driven approach involves relying on actual observational data, such as magnitude or flux values, rather than theoretical template SEDs. Specifically, this research aims to evaluate the reliability of Fully Connected Neural Networks (FCN) in estimating photometric redshift using magnitude data in light of recent advancements in the field of machine learning. Fully Connected Neural Networks, a subset of artificial neural networks, are designed to capture complex relationships in data, increasing overall predictive abilities for a model \cite{b9}. 

Despite the clear capabilities for neural network applications in astronomy, there remains a gap in comprehensive studies that use magnitude and color index data to make redshift predictions. Our research seeks to bridge this gap by comparing the Fully Connected Network with a decision tree regressor to see the efficacy of both when provided with light data from the SDSS for photometric redshift estimation. Although techniques like SVMs and random forests had similar scores when compared to the results of the decision tree regressor, we chose to use the decision tree regressor to compare against the neural network due to its use in past research on similar topics. The overall scope encompasses these models' design, training, and testing, followed by an analysis of their performance. Comparison metrics between the two methods will be RMS values and overall prediction accuracy.

\section{Methodology}

\subsection{Data}

Our study utilized a dataset from the Sloan Digital Sky Survey \cite{b10} with 50,000 celestial objects. For each of the objects, magnitudes of 5 different bands were included in the data. The 5 bands - u,g,r,i,z - represent different wavelengths of light from each galaxy or quasar \cite{b11}. Alongside the magnitudes, the dataset came with redshift value labels for each object. These redshifts were obtained from spectroscopic measurements from SDSS and the first 5 rows of data are shown (Table 1). 

\begin{table}[!h]
\caption{SDSS Dataset: Contains the first five rows of our original dataset for the object’s magnitude in different wavelength filters (u, g, r, i, and z) and the redshift values.}
\begin{center}
\begin{tabular}{|c|c|c|c|c|c|}
\hline
u & g & r & i & z & redshift\\
\hline
18.27449 & 17.01069 & 16.39594 & 16.0505 & 15.79158 & 0.0369225\\
18.51085 & 17.42787 & 16.94735 & 16.61756 & 16.46231 & 0.06583611\\
18.86066 & 17.91374 & 17.56237 & 17.26353 & 17.13068 & 0.1202669\\
19.38744 & 18.37505 & 17.63306 & 17.25172 & 17.00577 & 0.1806593\\
18.38328 & 16.59322 & 15.77696 & 15.3979 & 15.08755 & 0.04035749\\
\hline
\end{tabular}
\label{tab1}
\end{center}
\end{table}

Before diving into model training and testing, we visualized different portions of the data to better understand the distributions (Fig. 1). The pair plot showed that all of the bandpass filters were linearly dependent on one another as seen by the strong linear correlation between all the different filters and that they could have modeled some of the relations the network did when using these filters to predict redshift. 

\begin{figure}[ht!]
\centering
\includegraphics[scale = 0.45]{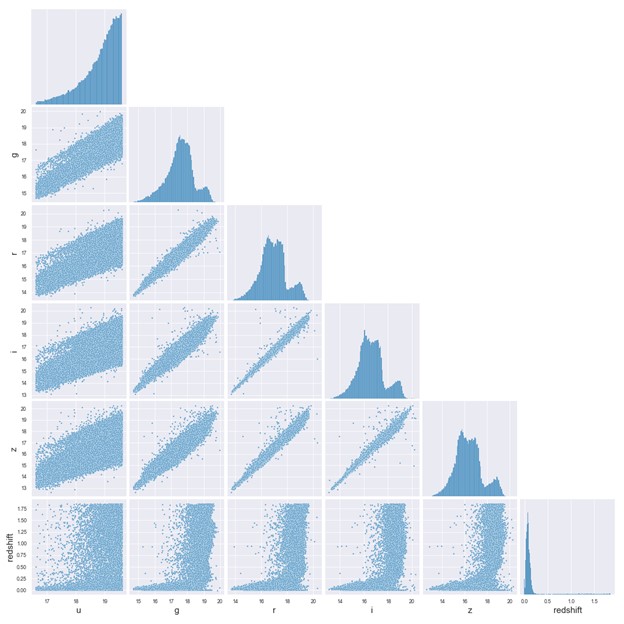}
\caption{Pair Plot for SDSS Data: The pair plot above shows the correlation between different magnitudes and their redshift values in the dataset. Each respective feature is paired with another and results in a scatterplot displaying the relationship between them. }
\label{fig1:magnitude}
\end{figure}

\subsection{Preprocessing}

We performed sigma-clipping on the redshift values using a sigma value of 3 standard deviations from the mean of the redshift values to remove outliers while retaining 95 percent of the data. The equation for the standard deviation is

\begin{equation}
	\sigma = \sqrt{ \frac{\sum{(x_i - \mu)}^2}{N}},
\end{equation}

Where $\mu$ is the mean of all the redshift values in the dataset, N is the total number of redshift values, and x represents each individual redshift value. 

Additionally, we removed redshift values less than zero as these are not physical. As a result, we ended with a dataset of 47,484 celestial objects out of the original 50,000. We justify this loss of data with the empirical rule that 95\% should be in between 3 standard deviations of the mean. Outliers and unnatural values initially caused the data to be skewed and thus made our model more biased and inaccurate. After preprocessing, we noticed that these methods helped a lot when making predictions by generalizing the data, stopping the overfitting of outliers, and making predictions less biased and more accurate.

\subsection{Decision Tree Regressor}

We compared two methods, a decision tree regressor and a fully connected neural network. The decision tree regressor (regression tree) works by partitioning the datasets into small subsets. Each split is based on the value of the input features. Our features consisted of the 5 bandpass filters (u,g,r,i,z) as well as the colors formed by their magnitude differences (u-g),(g-r),(r-i), and (i-z). After splitting the data, we arrive at leaf nodes where the redshift values are as similar as possible. Each leaf of the tree then predicts the average redshift of the instances that fall into it. The model is simple and transparent.

\subsection{Fully Connected Neural Network}

Our next step was to create model that accurately predicted redshifts, based on the given inputs. We developed a fully connected neural network that used the Adaptive Moment Estimation Optimizer \cite{b12} as it outperformed other models that we tested, such as random forests and multivariate regression models. However, fully connected neural networks do require many computational resources, especially when dealing with larger datasets. In our study, we used powerful computers with faster GPUs, yet still had to give the model time to run through many epochs. For even larger datasets, the model would need to be optimized to efficiently handle the computational demands. 

Our input layer consists of 9 inputs and an output shape of 100. The 9 inputs are composed of magnitudes across each of the band passes and the magnitude differences (colors) as follows, 

\begin{equation}
    {M}_{Input} = [m_u, m_g, m_r, m_i, m_z, m_{u-g}, m_{g-r}, m_{r-i}, m_{i-z}].
\end{equation}

When tuning our model, we used GridSearchCV from sci-kit-learn to find the best number of neurons in each layer and this algorithm also selected 9 inputs (5 bandpass filters and the 4 colors) instead of just the bandpass filters to create an optimal model. Although the neural network still performed well with just the original bandpass filters, adding the colors as features decreased RMS further and was in line with the results from the GridSearchCV model tuning. This tuning showed that the network is robust as it can take different combinations of parameters and still perform well when predicting redshift. Therefore, we added two more hidden layers of neurons before the last layer with 65 neutrons and 35 neurons, respectively, instead of following the standard architecture in DNNs. With only one neuron in this last layer, it represents the predicted redshift. We used the Rectified Linear Unit (ReLU) activation function \cite{b13} which worked better than the sigmoid function to account for redshift predictions with values greater than 1 as well as to improve the efficiency of the network. Further, we used the Adaptive Moment Estimation (ADAM) Optimizer because of its faster convergence than stochastic gradient descent, its ability to bypass local minimums to find the global minimum, and its overall effectiveness. It has these advantages by keeping an exponentially decaying average of past gradients and squared gradients. Further, we tried stochastic gradient descent and RMSprop for optimizers and found that ADAM performed the best. Lastly, we added a dropout rate of 0.2 to prevent overfitting after each layer which finalized the composition of the neural network’s architecture (Fig. 2). 

\begin{figure}[ht!]
\centering
\includegraphics[scale = 0.6]{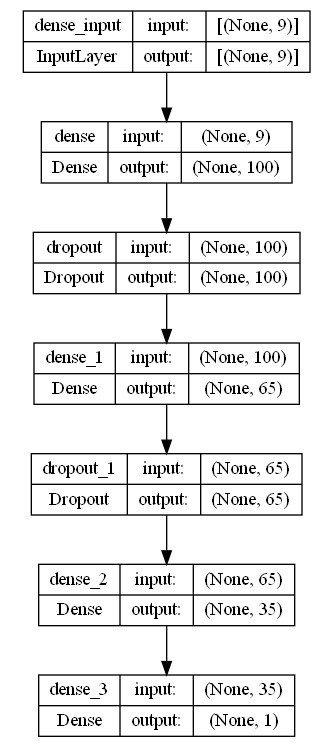}
\caption{Neural Network Architecture: The chart above shows the layers and dropouts for the neural network and how the model’s neurons are arranged to obtain the redshift value from the nine input parameters. }
\label{fig2:magnitude}
\end{figure}

\subsection{Loss Function - Mean Squared Error}

We minimize the mean squared error as the loss function in our neural network, 

\begin{equation}
	\mathcal{L} = \frac{1}{n} \sum_{i=1}^{n}{(y - \hat{y})^2},
\end{equation}

where y is the true redshift, y-hat is the predicted redshift, and n is the number of objects in a batch of the training set. 

\subsection{Validation Data}

To test the performance of our model, we used SKLearn to split the data into a training set with 80 percent of the data and a testing set with the remaining 20 percent of the data. Before making the split, we shuffled the data to randomize it to get comparable data in both the training and testing sets. 

\section{Results}

For the original decision tree regressor, the RMS value was above 0.16. In addition, when graphing the true redshift values versus the predicted values given by the tree regressor, we see a graph with a lot of noise and many values far from the predicted line (Fig. 3). Using our neural network, we were able to improve accuracy when predicting redshift to 0.009 RMS using the predicted line (y-hat) that our neural network produced (Fig. 4). It's also important to note that the scale in figure 4 is smaller (with a true redshift range of 1.3) than the true redshift range in figure 3 (with a true redshift range of 2) because we used a smaller test split when using train\_test\_split method in scikit learn.

The improvement in the RMS show that the neural network predictions have improved upon those from the Decision Tree Model (previously used to predict photometric redshift in stars and quasars) due to a lower RMS and lower deviations and discrepancies from predicted redshift to actual redshift as observed in the graphs above (Fig. 4). Moreover, the neural network also had a better mean absolute error (MAE) with fewer false positives than the decision tree regressor. 

In addition, the model was trained to prevent overfitting to our dataset, so our results can be generalized to any data in this particular data format, given all nine parameters are present (Fig. 5).

\begin{figure}[ht!]
\centering
\includegraphics[scale = 0.4]{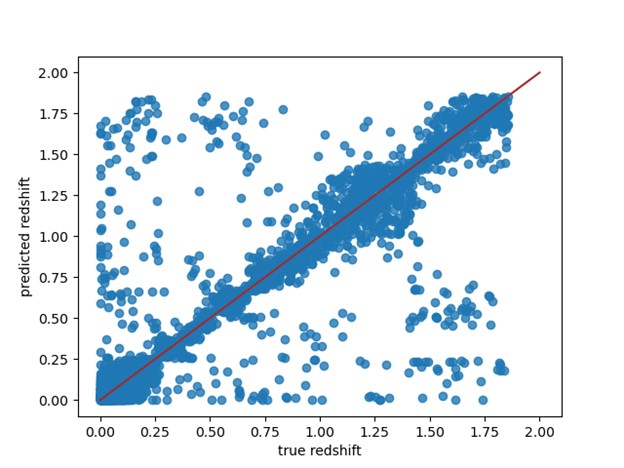}
\caption{Decision Tree Regressor Scatter Plot: The graph shows true redshift vs predicted redshift correlation for the predictions made by the decision tree regressor. }
\label{fig3:magnitude}
\end{figure}

\begin{figure}[ht!]
\centering
\includegraphics[scale = 0.58]{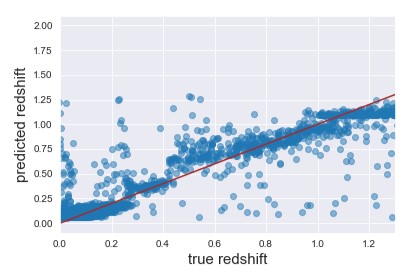}
\caption{Fully Connected Neural Network Scatter Plot: The graph above shows true redshift vs predicted redshift correlation for the fully connected neural network’s predictions. There are few outliers, with most predictions being close to the best fit line.}
\label{fig4:magnitude}
\end{figure}

\begin{figure}[ht!]
\centering
\includegraphics[scale = 0.58]{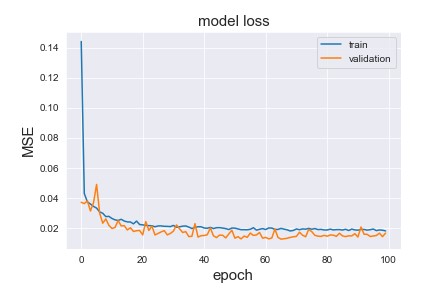}
\caption{Line Graph showing Epochs vs MSE: This graph shows the learning curve of the fully connected neural network. The loss of the training function followed the same trajectory as that of the validation set, stabilizing and reaching an equilibrium, indicating a good fit.}
\label{fig5:magnitude}
\end{figure}

\newpage\section{Results Analysis \& Implications}
The implications of the study extend beyond the test dataset. The empirical evidence from our study has not only demonstrated a data-driven approach but has shed light on incredibly efficient methods for photometric redshift estimation. The refined precision of redshift values delivered by our FCN is poised to enhance the identification process, ensuring more accurate categorizations for distance estimation of objects. 

The strength of our FCN lies in its generalizability and usability. Since it thrives on capturing intricate data relationships, the model can adapt to datasets with varying structures, enhancing its scalability and overall robustness. Furthermore, this broadens the scope of its application, making it suitable for different datasets.

This shift towards a data-driven methodology holds strong implications for the future of astronomical research. Compared to the traditional decision tree regression models, the FCN showcases a clear edge in estimating redshifts. This distinction becomes paramount when differentiating between quasars and stars. Moreover, as the acquisition of SED templates becomes increasingly challenging, the need for approaches that can efficiently utilize raw astronomical data will become more pressing. Our FCN model, with its strong adaptability and scalability, stands as a testament to the possibilities that lie in store for data-driven astronomy. 

Yet, our study does have certain limitations. Despite the vastness of our dataset, 50,000 celestial objects might still just be scratching the surface. The universe's expanse and the inherent variability within it mean that larger datasets could exhibit different behaviors, and while our FCN is promising, its real test would be in even more diversified astronomical conditions. Additionally, similar to most neural networks, FCNs are often seen as black box models, limiting their interpretability. Future iterations of our model could benefit from incorporating explainability techniques like SHAP (SHapley Additive exPlanations), which reveals the most influential features driving predictions within the model. Furthermore, while our model was trained to prevent overfitting, the complexity of FCNs may result in sensitivity to subtle variations within the training data. Future work could implement techniques like the Monte Carlo dropout to provide greater confidence within our predicions.

\section{Conclusion}
Overall, our model is very robust and its predictive capability surpasses the TPZ [6] approach, the artificial neural network [7], and the decision tree regressor used in this paper. The best application of our model would be to integrate it with the aforementioned past methods through heterogeneous ensembling and stacking, which would yield even more accurate redshift predictions. Additionally, our approach can be used alongside traditional template-based methods to further improve the precision of photometric redshift estimation. Instead of solely relying on an FCN for redshift predictions, combining our methodology with existing models would result in better insights.

\section{Future Work}
Building on the study’s foundation, there are various options for improvement and future research. Given the success of the FCN, the integration of hybrid models, such as combining FCNs with Random Forests could improve overall accuracy. One specific approach is using heterogeneous ensemble learning using a stacking approach. This involves tree-based and KNN models to make predictions before feeding those predictions to a decision boundary-based algorithm(such as logistic regression or SVM) to come up with a final model for estimating redshift.  Lastly, fine-tuning pre-trained models would build on the foundation of previous successes could possibly improve predictive power for redshift.

\section*{Acknowledgment}
Although we are currently freshmen in university, most of this work was conducted in collaboration with the Cambridge Centre for International Research (CCIR) during our senior year of high school. We extend our appreciation to CCIR for their invaluable contributions and resources throughout this project.

Additionally, we would like to express our deep gratitude to Dr. Daniel Muthukrishna and Fatima Zaidouni from the MIT Kavli Institute for Astrophysics and Space Research for their guidance and support as our senior supervisor and mentor, respectively.

\vspace{12pt}

\end{document}